\documentclass[superscriptaddress,aps,prx,twocolumn,showpacs,nofootinbib, longbibliography, notitlepage,floatfix]{revtex4-2}
\usepackage{amsmath,amsfonts,amssymb,amsthm,graphics,graphicx,epsfig,bbm}
\usepackage[colorlinks=true,citecolor=blue,linkcolor=blue,urlcolor=blue]{hyperref}
\usepackage[usenames]{color}
\usepackage{graphicx}
\usepackage{subfigure}
\usepackage{amsmath}
\usepackage{epsfig}
\usepackage{dcolumn}
\usepackage{bm}
\usepackage{color}
\usepackage{times}
\usepackage{epstopdf}
\usepackage{soul}
\usepackage{amssymb}
\usepackage{amstext}
\usepackage{latexsym}
\usepackage{hyperref}
\usepackage{amsfonts}
\usepackage{psfrag}
\usepackage{soul,xcolor}
\usepackage[normalem]{ulem}
\usepackage{dsfont}
\usepackage{txfonts}
\usepackage{physics}
\usepackage{xcolor}
\usepackage{footnote}
\usepackage{multirow}
\usepackage{appendix}
\usepackage{mathtools}
\usepackage{ulem}
\usepackage{mathtools}

\newtheorem{thm}{Theorem}

\theoremstyle{definition}

\theoremstyle{remark}

\usepackage{todonotes}

\usepackage{mathtools}
\DeclarePairedDelimiter{\floor}{\lfloor}{\rfloor}

\begin{document}

\setstcolor{red}
\newtheorem{Proposition}{Proposition}[section]	

\title{Efficient Hamiltonian Simulation for Solving Option Price Dynamics} 
\date{\today}

\author{Javier Gonzalez-Conde}
\email[Corresponding author: ]{\qquad javier.gonzalezc@ehu.eus}
\affiliation{Department of Physical Chemistry, University of the Basque Country UPV/EHU, Apartado 644, 48080 Bilbao, Spain}
\affiliation{EHU Quantum Center, University of the Basque Country UPV/EHU, Apartado 644, 48080 Bilbao, Spain}

\author{\'Angel Rodr\'iguez-Rozas}
\affiliation{Risk Division, Banco Santander, Avenida de Cantabria S/N, 28660 Boadilla del Monte, Madrid, Spain}

\author{Enrique Solano}
\affiliation{Kipu Quantum, Greifswalderstrasse 226, 10405 Berlin, Germany}

\author{Mikel Sanz}
\email[Corresponding author: ]{\qquad mikel.sanz@ehu.eus}
\affiliation{Department of Physical Chemistry, University of the Basque Country UPV/EHU, Apartado 644, 48080 Bilbao, Spain}
\affiliation{EHU Quantum Center, University of the Basque Country UPV/EHU, Apartado 644, 48080 Bilbao, Spain}
\affiliation{IKERBASQUE, Basque Foundation for Science, Plaza Euskadi 5, 48009, Bilbao, Spain}
\affiliation{Basque Center for Applied Mathematics (BCAM), Mazarredo Zumarkalea, 14, 48009 Bilbao, Spain}

\begin{abstract}

Pricing financial derivatives, in particular European-style options at different time-maturities and strikes, means a relevant problem in finance. The dynamics describing the price of vanilla options when constant volatilities and interest rates are assumed, is governed by the Black-Scholes model, a linear parabolic partial differential equation with terminal value given by the pay-off of the option contract and no additional boundary conditions. Here, we present a digital quantum algorithm to solve Black-Scholes equation on a quantum computer by mapping it to the Schr\"odinger equation. The non-Hermitian nature of the resulting Hamiltonian is solved by embedding its propagator into an enlarged Hilbert space by using only one additional ancillary qubit. Moreover, due to the choice of periodic boundary conditions, given by the definition of the discretized momentum operator, we duplicate the initial condition, which substantially improves the stability and performance of the protocol. The algorithm shows a feasible approach for using efficient Hamiltonian simulation techniques as Quantum Signal Processing to solve the price dynamics of financial derivatives on a digital quantum computer. Our approach differs from those based on Monte Carlo integration, exclusively focused on sampling the solution assuming the dynamics is known. We report expected accuracy levels comparable to classical numerical algorithms by using 9 qubits to simulate its dynamics on a fault-tolerant quantum computer, and an expected success probability of the post-selection procedure due to the embedding protocol above 60\%.
\end{abstract}

\maketitle

\newtheorem{theorem}{Theorem}[section]
\newtheorem{corollary}{Corollary}[theorem]
\newtheorem{lemma}[theorem]{Lemma}
\def\endproof{\hfill$\blacksquare$}

\section{Introduction}

In finance, European-style vanilla options are financial derivative contracts written on an underlying asset, which give the holder the right to buy or sell such asset on a specified future date at a predetermined strike price. One of the fundamental tasks of quantitative finance is to calculate a \textit{fair price} of such option contract before its expiration time. This task is far from being straightforward due to the randomness associated to the time evolution of both the underlying stock and the interest rates, whose dynamics can be modelled via either a stochastic processes or a partial differential equation (PDE), both connected by the celebrated Feynman-Kac formula. One of the first successful approaches to this problem was achieved by F. Black and M. Scholes in $1972$, who proposed the celebrated Black-Scholes model \cite{FBMS}, in which a lognormal distribution of the underlying stock price is assumed. Even though a closed-form solution exists for this dynamics, the numerical method proposed in this manuscript relies on the Black-Scholes model in order to show its properties of convergence and accuracy. Moreover, we show that this method is also applicable to time-dependent volatility PDEs, for which non closed-form solution exists in general. Additionally, the scope of this paper is to present a novel algorithm thought to be extended to more complex PDEs as a future work. Besides, numerical solutions also turn out to be fundamental when hedging a portfolio with a great number of coupled options. Several classical methods proposed in the literature include finite differences, finite elements, Monte Carlo methods, and Fourier spectral methods \cite{FVM,MEV,MPAMC,WFDM,WFDM2}.

Quantum technologies have undergone rapid development in the last decade, paving the way for transformative advancements in various fields. Quantum technologies have experienced a rapid development in the last decade. Recently, Google has achieved quantum advantage, meaning that they have performed a calculation employing a superconducting processor faster than the most powerful supercomputers available today \cite{GOOGLE}. Among these domains, finance is poised to experience a profound impact from this emerging technology. Indeed, the emergence of scalable quantum technologies will affect forecasting, pricing and data science, and will undoubtedly have an economic impact in the following years \cite{QFFP, OP}. Certainly, there already exist several efforts in this direction, for instance, an attempt to predict financial crashes \cite{PFC,PFCDW}, the application of the principal component analysis to interest-rate correlation matrices \cite{PCAS},  quantum methods for portfolio optimization \cite{RQA, QAPO,DPO,PO40,PO60,han2022quantum}, quantum generative models for finance \cite{GMF}, a quantum model for pricing collateral debt obligations \cite{CDO}, a protocol to optimize the exchange of securities and cash between parties \cite{QATS} and an application to improve Monte Carlo methods in risk analysis \cite{QRA,CRA,matsakos2023quantum}, among many others. 

Regarding the option pricing problem, it has been studied the problem of sampling the solution resulting from the stochastic process of the Black-Scholes model by employing Monte Carlo methods, and assuming its dynamics is known at any maturity time. In Ref.~\cite{QMC}, the authors proposed a theoretical approach to sample the solution of the stochastic process using quantum Monte Carlo integration, reporting a quadratic speedup versus classical sampling techniques. Afterwards, an experimental implementation in the IBM Tokyo quantum processor was attained in Refs.~\cite{PQC,IQAE,TQA}, employing a gate-based methodology to price options and portfolios of options. Additionally, several approaches to solve the stochastic process were proposed in \cite{QUA,Alghassi2022variationalquantum, An_2021, li2023quantum,rebentrost2022quantum}. Relevant alternative perspectives to deal with pricing problems involving linear partial differential equations consist in adapting quantum algorithms applied to existing quantum numerical solvers \cite{FEM,HPA,LDE,Liu_2021, Berry_2014, navier, Krovi_2023, leyton2008quantum, an2022efficient,https://doi.org/10.4230/lipics.tqc.2022.2}, or even use variational and generative approaches  \cite{zhao2022quantuminspired,certo2023conditional, kubo2022pricing}.

In this Article, we propose a quantum algorithm for solving the dynamics of the Black-Scholes partial differential equation on a quantum computer based on Hamiltonian simulation techniques  \cite{Lloyd_1996,Preskill_2018, Tacchino_2019, Georgescu_2014, Barends_2016, Camps_2022, Trotter_1959, Clinton_2021, Efekan_2022, SSH, Low_2019, Childs2012, Suzuki_1976, Pastori_2022, Dong_2021, Campbell_2019, Layden2022, Watkins_2022, Berry_2020, Chen_2021, Berry_2015, jin2022quantum, jin2023quantum,Low_2017}. In this sense, our manuscript focuses on the Hamiltonian simulation step as a subroutine to solve the option pricing problem and does not aim to discuss the end-to-end process of such a task. In order to achieve this practical task in an efficient way, one can follow the methodology described in \cite{Miyamoto_Kubo}, where the pricing is obtained by solving the backwards PDE up to not the
present but some future time. In this way, it is possible to evolve the price curve backwards, which we illustrate in this manuscript via Hamiltonian simulation, and the underlying stock forwards (either via SDE or PDE). Finally it is possible to efficiently compute the expected value of the price under the distribution of the stock for obtaining the pricing of the derivative. To this end, we map Black-Scholes equation into the Schr\"odinger equation, which results into an equivalent problem consisting of simulating a non-Hermitian Hamiltonian. Additionally, we impose periodic boundary conditions to achieve an efficient diagonalization of the discretized momentum operator into a quantum computer via the discrete Fourier transform. In order to simulate the non-unitary dynamics into a quantum processor \cite{QS,FQS,NODP,SWE,SWEQA,PTSYM}, we embed the time propagator into an enlarged Hilbert space, making use of only one ancillary qubit, using a technique known as \textit{unitary dilation}. Thanks to this embedding we can post-select the result depending on the outcome of the ancillary qubit, which allows us to reproduce the dynamics of the non-Hermitian Black-Scholes Hamiltonian. Moreover,  we use one of the qubits of the spatial discretization to duplicate the initial boundary condition to fit the periodic boundary conditions, leading to an improvement of the performance, accuracy and stability of the algorithm by mitigating edge effects propagation. In comparison with previous methods, our algorithm presents a Hamiltonian simulation methodology to solve the Black-Scholes partial differential equation instead of solving the stochastic differential equation. The simulations show a precision comparable to classical algorithms with a quantum circuit comprising 9 qubits to simulate the dynamics of the PDE in a fault-tolerant quantum computer, and furthermore, an expected success probability of the post-selection protocol above 60\%. 

The article is structured as follows. First, we briefly review the Black-Scholes model and map it to a Hamiltonian formulation. We propose an embedding protocol and present the digitalization of the space used to encode the problem into a digital quantum computer. Next, we provide the details of our algorithm and depict its circuit implementation. Finally, we show the results and discuss the future scopes.

\section{Black-Scholes Schr\"odinger Equation}
\label{sec2}
Under the assumption of constant interest rate and volatility, and provided certain ideal market conditions, Black-Scholes model \cite{FBMS} is based on the possibility of building a perfect dynamic  hedging portfolio strategy, known as \textit{delta hedging}, which consists in holding, at each time, a number of shares equal to the derivative of the option price with respect to stock price. Therefore, the only risky (random) factor associated to portfolio dynamics is eliminated and the value of the portfolio agrees with the option value at any time. The pricing problem for a specific derivative contract, i.e. to determinate its present price $V(t=0,S)$, is given by the Black-Scholes PDE,

\begin{equation}
\frac{\partial V}{\partial t}+ rS\frac{\partial V}{\partial S}+\frac{1}{2}\sigma^2S^2\frac{\partial^2 V}{\partial S^2}=rV,
\label{BS}
\end{equation}
together with the terminal condition for the price of the option given by the pay-off of the option contract, $V(t=T, S)$, defined at maturity time, $T$, for any plausible value of the underlying stock $S \geq 0$   and on the strike price, $K$. Here, $r$ represents the constant risk-free interest rate while $\sigma$ is the constant volatility of the stock, both assumed to be constant. In the case of a European Put-type option, the pay-off function reads $V_p(T, S)=\text{max}\{K-S,0\}$. Typical solutions to these PDE are shown in Fig. \ref{Diser}.

\begin{figure}[b]
\centering
\includegraphics[width=0.5\textwidth]{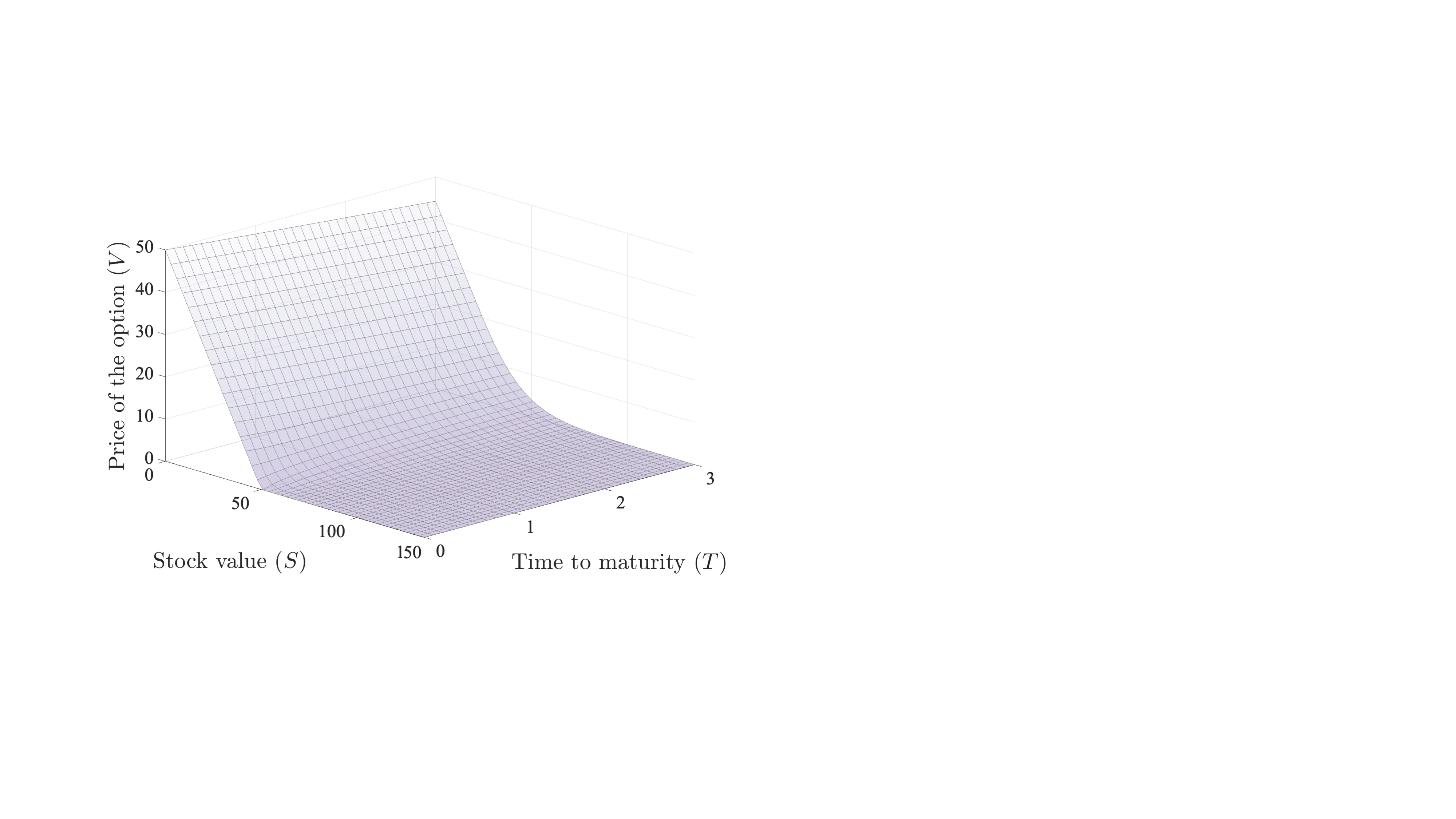}
\caption{Typical solutions of Eq.~(\ref{BS}) for a European Put-type option. Simulation parameters: $S_\text{max}=150 $ u, $K=50 $ u, $\sigma=0.2$ and $r=0.04$.}
\label{Diser}
\end{figure}

The Black-Scholes equation has a similar structure to Schrödinger equation \cite{QPF}, which suggests the possibility of efficiently simulating such model on a quantum platform. To that end, we rewrite the Black-Scholes equation in a Hamiltonian form. First, the change of variables $ S=e^x$, $-\infty < x < \infty $, allows us to recover the unbounded position variable, leading to the equation

\begin{equation}
    \frac{\partial V}{\partial t}+\left(r-\frac{\sigma^2}{2}\right)\frac{\partial V}{\partial x}+\frac{\sigma^2}{2}\frac{\partial^2 V}{\partial x^2}=rV.
    \label{BSlog}
\end{equation}
Note that this equation is a backward parabolic equation.  Thus we can reverse time $t \rightarrow \tau = T-t$, obtaining a forward parabolic equation, and consequently an initial value problem where $V_p(\tau=0, S)=\text{max}\{K-S,0\}.$\\

Finally, let us also introduce the momentum operator $\hat{p}:=~-~i\frac{\partial}{\partial x}$ to rewrite Eq.~(\ref{BS}) as

\begin{equation}
 \frac{\partial V}{\partial \tau}=i\hat{H}_{BS}V
\label{SEQ}
\end{equation}
where we define
\begin{equation}
\label{QBS}
\hat{H}_{\text{BS}}=i\frac{\sigma^2}{2}\hat{p}^2-\left(\frac{\sigma^2}{2}-r \right)\hat{p}+ir\mathds{I}.
\end{equation} 
as the Black-Scholes Hamiltonian. Solutions to Eq.~(\ref{SEQ}) are given by the time propagator $\hat{U}(\tau)=e^{i\tau\hat{H}_{BS}}$ acting on the initial condition. We can observe in Eq.~(\ref{QBS}) that the Black-Scholes Hamiltonian is a non-Hermitian operator, i.e., $\hat{H}_{BS}\neq \hat{H}_{BS}^{\dagger}$, which implies that neither its eigenvalues are necessarily real, nor the associated  time propagator, $\hat{U}(\tau)$, is unitary.The evolution of a closed quantum system is always unitary. However, this poses a significant challenge when trying to find a physical system that follows the dynamics of the Black-Scholes model. To address this issue, we introduce a technique in Section \ref{embedding} where we embed the propagator into a larger space by utilizing an ancillary qubit. Subsequently, in Section \ref{postselection}, a post-selection technique is employed to effectively retrieve the desired Black-Scholes dynamics.

Therefore, by presenting the efficient mapping from the partial differential equation (PDE) of the Black-Scholes equation to its Hamiltonian, we have established that the complexity of solving the Black-Scholes equation is equivalent to that of the non-Hermitian Hamiltonian simulation problem. Note that alternatively there exists transformations that map the Black Scholes equation to a heat equation \cite{heateq_book, Alghassi2022variationalquantum}, nevertheless this does not solve the non Hermitian nature of the Hamiltonian.

\subsection{Embedding Protocol}
\label{embedding}
The Black-Scholes Hamiltonian, Eq.~(\ref{QBS}),  can be decomposed into a Hermitian and an anti-Hermitian part, i.e. $\hat{H}_{\text{BS}}=~\hat{H}_{\text{BSH}}+\hat{H}_{\text{BSA}}$, with

\begin{equation}
\label{bseq}
\hat{H}_{\text{BSH}}=-\left(\frac{\sigma^2}{2}-r\right)\hat{p}, \ \ \ \ \hat{H}_{\text{BSA}}=i\left(\frac{\sigma^2}{2}\hat{p}^2+r\mathds{I}\right).
\end{equation}
Additionally, we have that $[\hat{H}_{\text{BSH}},\hat{H}_{\text{BSA}}]=0$, so via the Baker–Campbell–Hausdorff formula \cite{BCH}, the propagator can be written as  $\hat{U}(t)=e^{i\tau\hat{H}_{\text{BSA}}} \ e^{i\tau\hat{H}_{\text{BSH}}}$. Furthermore, notice that $\hat{O}(\tau)=e^{i\tau\hat{H}_{\text{BSA}}}$ is an Hermitian operator. \\

In order to circumvent the problem of dealing with the non-Hermitian operator, we embed the propagator $\hat{O}(\tau)$ into a larger unitary operator using a technique from operator theory called \textit{unitary dilation} \cite{UD}. Indeed, by adding an ancillary qubit, $q_E$, to our system, we can embed $\hat{O}(\tau)$ into the unitary operator $\tilde{U}(\tau)$ which can be written as 

\begin{equation}
\tilde{U}(\tau)=\begin{pmatrix}
\hat{O} & \sqrt{1-\hat{O}^2} \\
\sqrt{1-\hat{O}^2} & -\hat{O}
\end{pmatrix}=
\left(\hat{\sigma}^z_E\otimes \mathds{I} \right)\exp \left(i\hat{\sigma}^y_E\otimes\tilde{H}(\tau)\right),
\label{Uemb}
\end{equation}
with $\tilde{H}(\tau)= \text{arccos}{(\hat{O}(\tau))}$ the `integrated embedded Hamiltonian' and $\norm{\hat{O}(\tau)}_2\leq 1$, with $||\cdot ||_2$ the spectral norm. In case $ \norm{\hat{O}(\tau)}_2> 1$ one just have to renormalize the operator. In our particular case as all the eigenvalues of $\frac{\sigma^2}{2}\hat{p}^2+r\mathds{I}$ are positive, then the spectral norm of the exponential $e^{-(\frac{\sigma^2}{2}\hat{p}^2+r\mathds{I})}$ is smaller than 1.
\\

Starting from the initial state $\ket{\Phi_0}=\ket{0_E}\otimes \ket{V_p}$, with $\ket{V_p}$ encoding the pay-off condition of the European Put-type option, the system evolves according the unitary operator $\tilde{U}(t)$ to obtain the final state 
\begin{equation}
\ket{\Phi}=\hat{O}\ket{0_E}\otimes \ket{V_p} + \sqrt{1-\hat{O}^2}\ket{1_E}\otimes \ket{V_p}.
\end{equation}
If we apply a post-selection technique filtering the outcomes with the ancillary qubit in the state $\ket{0_E}$, we can simulate the propagator $\hat{O}(\tau)$ into a quantum computer, and in consequence, the whole Black-Scholes Hamiltonian dynamics. We provide the details of the state preparation and post-selection process in Section \ref{VE}.

Note that this methodology based on the \textit{unitary dilation} only introduces an additional qubit as an extra computational resource, so we consider it does not increase the complexity class of the non-Hermitian Hamiltonian simulation problem.

\subsection{Digitization of the space}

In order to perform a digital simulation of the Black-Scholes equation using a quantum computer, it is required a discretization of position and momentum spaces based on the number of qubits employed. The possibility of simulating the Black-Scholes model on a discretized space is guaranteed by the Nyquist-Shannon sampling theorem \cite{Shannon}.  Following the work  in Ref. \cite{DSF}, a wave function $\ket{\Psi}$ such that $|\Psi(x)| < \epsilon$ when $|x| > x_\text{max}$  and whose Fourier transform  $|\hat{\Psi}(p)|<\epsilon$  if  $|p| \geq x_\text{max} $ can be sampled in position space using the basis of sampling vectors $\{\ket{x_j}\}$ where $x_j = -x_\text{max}+~ j\delta_x$, with   $\delta_x\leq\frac{\pi}{x_\text{max}}$ and $j =~ 0,1,...,N_x-1$ such that $x_j\in ~[-x_\text{max},x_\text{max}]$. For a given interval, in the limit where $\delta_x=\frac{\pi}{x_\text{max}}$, the minimum $N_x$ is given by the equality $2x_\text{max}=~\delta_x (N_x-~1)$. Hence, the wave function can then be rewritten as $\ket{\Psi}=~\sum_{j=0}^{N_x-1}\Psi(x_j)\ket{x_j}$.   The conjugate momentum basis is obtained by the discrete Fourier transform of the position basis, $\text{QFT}:~\ket{x_j} \mapsto~\ket{p_j} =~ \frac{1}{\sqrt{N_x}} \sum_{k=0}^{N_x-1} \omega_{N_x}^{jk} \ket{x_k}$, where $\omega_{N_x}= e^{\frac{2 \pi i}{N_x}}$. We denote the discrete quantum Fourier transform matrix operator as $\hat{F}$. These two sampling basis allow us to define the following discretized position and momentum operators acting on their own basis as  $\hat{X}_x\ket{x_j}=~x_j\ket{x_j}$ and  $\hat{P}_k\ket{p_k}=p_k\ket{p_k}$.

We consider an equispaced grid of the interval $[-x_\text{max},x_\text{max}]$, thus, the position space is discretized into  the values $x~=~-~x_\text{max}~+~\delta_x ~\beta_x$, with $\delta_x=\frac{2x_\text{max}}{N_x-1}$ and $\beta_x=0, \hdots , \ N_x-1$. If we consider that the position $x= - x_\text{max}$ is represented by the state $\ket{- x_\text{max}}=\ket{0...0}$, and the position $x=x_\text{max}$ is represented by $\ket{x_\text{max}}=\ket{1...1}$, the matrix form of this operator in the $x$ basis results

\begin{equation}
\centering
\hat{X}_x=x_\text{max}\begin{pmatrix}
-1 & 0 & \hdots & 0 & 0 \\
0 & -1+\delta_x & \hdots & 0 & 0 \\
\vdots & \vdots & \ddots & \vdots & \vdots \\
0 & 0 &  \hdots & 1-\delta_x  & 0 \\
0 & 0 & \hdots &  0 & 1 \\

\end{pmatrix}.
\end{equation}

Let us now construct the momentum operator, $\hat{p}$. By using the second order of finite differences, we approximate the derivative of a certain function as

\begin{equation}
   \frac{d \text{f}(x)}{d x} \approx \frac{\text{f}(x+\delta_x)-\text{f}(x-\delta_x)}{2\delta_x}.
\end{equation}
Consequently, imposing periodic boundary conditions, the discrete momentum operator in the position basis is given by the matrix

\begin{equation}
\centering
\hat{P}_x=\frac{-i}{2\delta_x}\begin{pmatrix}
0 & 1 & 0 & \hdots & 0 & \boxed{-1} \\
-1 & 0 & 1 & \hdots & 0 & 0 \\
\vdots & \vdots & \vdots & \ddots & \vdots & \vdots \\
0 & 0 &  \hdots & -1 & 0 & 1 \\
\boxed{1} & 0 & \hdots & 0 & -1 & 0 \\

\end{pmatrix}.
\label{TBC}
\end{equation}
Thanks to the choice of periodic boundary conditions, the momentum matrix $\hat{P}_x$ belongs to circulant matrix class, and therefore its diagonal form is obtained by using the discrete Fourier transform unitary matrix, $\hat{F}_{N_x}$,

\begin{equation}
    \hat{P}_k=\hat{F}_{N_x}\hat{P}_x \hat{F}_{N_x}^{\dagger}.
    \label{Fourier}
\end{equation}
This transformation can be efficiently implemented in a quantum computer \cite{CHUANG}. Otherwise we can not ensure the efficient diagonalization of the momentum matrix, incurring into an exponential cost in the general case. The analytical expression of the eigenvalues of $\hat{P}_x$ is also known, and is described by the equation
\begin{equation}
    p_k=\frac{\sin{\left(\frac{2\pi k}{N_x}\right)}}{\delta_x}, \ \ \ k=0 \ ... \ N_x-1.
    \label{eigen}
\end{equation}

Alternatively, the derivative operator can be defined according to the Discrete Fourier Transform definition, which implicitly assumes periodic boundary conditions.  In its own basis, the diagonal derivative operator, $\partial_x$, can be defined as

\begin{equation}
\centering
i k_\text{max} \begin{pmatrix}
-1 & 0 & \hdots & 0 & 0 \\
0 & -1+\delta_k & \hdots & 0 & 0 \\
\vdots & \vdots & \ddots & \vdots & \vdots \\
0 & 0 &  \hdots & 1-2\delta_k  & 0 \\
0 & 0 & \hdots &  0 & 1-\delta_k  \\
\end{pmatrix},
\label{TBC}
\end{equation}
with $k_\text{max}=\frac{\pi}{\delta_x}$, $\delta_k=\frac{2\pi}{2^n\delta x}$. The representation of this operator in the position space can be obtained by using the discrete Fourier transform unitary matrix.


\begin{figure}[t!]
\centering
\includegraphics[width=0.475\textwidth]{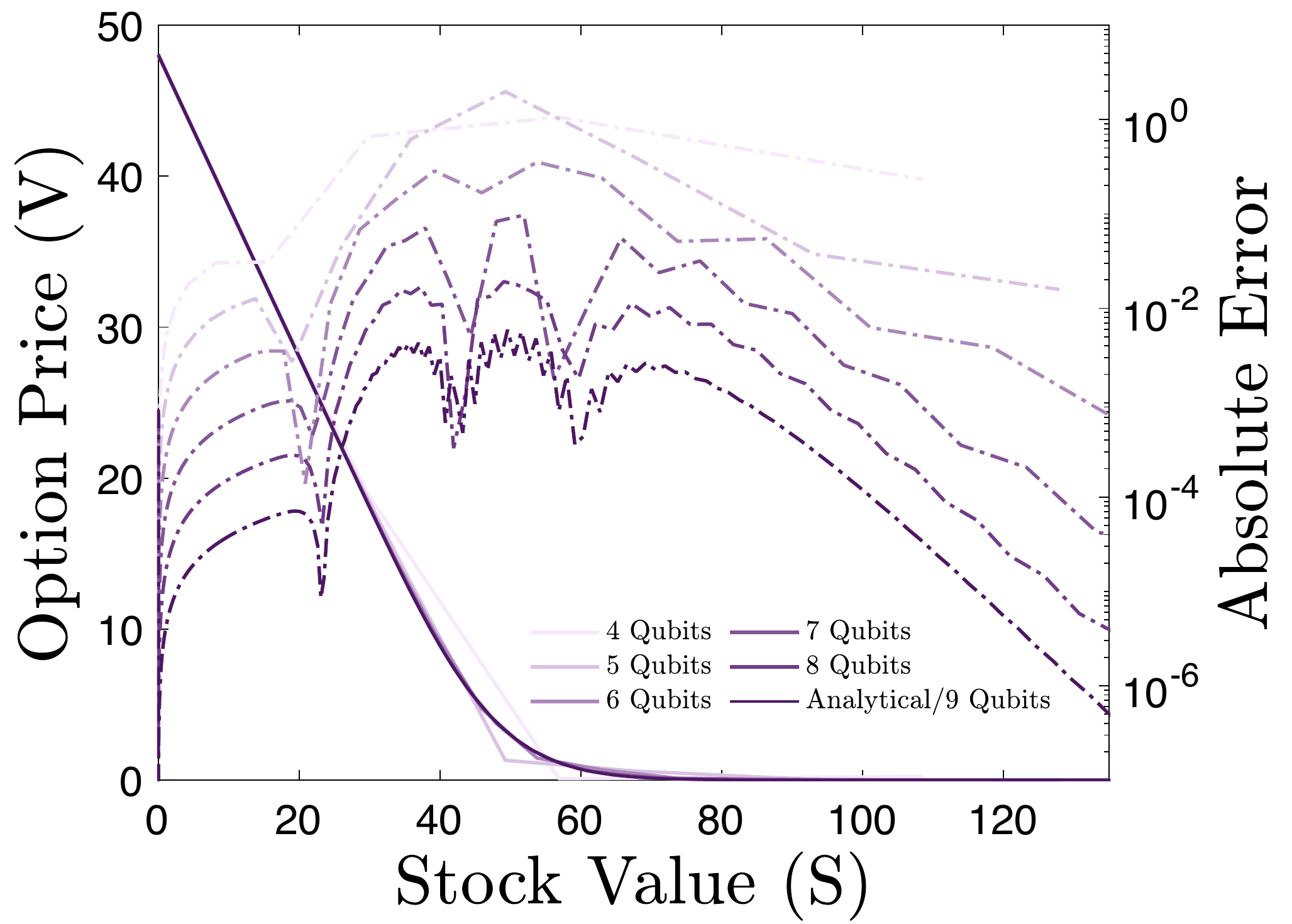}
\caption{Continuos line: convergence of the solution of Black-Scholes put option pricing problem obtained with the finite differences discretized operator $\hat{P}_x$ for distinct number of qubits, $n= 1\hdots 8$ (excluding the ancillary qubit used to duplicate the initial condition) and the analytical solution. Dashed line: discretization error per point depending on the number of qubits, $n= 1\hdots 9$ (excluding the ancillary qubit used to duplicate the initial condition). Simulation parameters: $S_\text{max}=150 $ u, $K=50 $ u, $\sigma=0.2$, $r=0.04$, $T=1$ year. Simulations have made use of the duplication of the initial condition. For source code of the simulations see \cite{github}}.
\label{Diser}
\end{figure}

In Fig. \ref{Diser}, we illustrate the convergence of the solution to Black-Scholes equation and its relative discretization error with respect to the analytical solution for different number of qubits obtained by making use of the momentum discrete operator given by Eq.~(\ref{TBC}) to simulate the Hamiltonian dynamics. For these simulations we have made use of the duplication of the initial condition that we explain in section \ref{VA}.

\section{Implementation on a Quantum Computer}
\label{VE}
In this section, we show the different subroutines of the circuit that simulates the price evolution for a put option contract in a quantum computer: information loading, Hamiltonian simulation and post-selection. The procedure for a call option would be similar but initializing the process in the corresponding pay-off state. In the following sections we will assume that $K\geq 1$ and the constraint $x_\text{max}=\log S_\text{max}=2\log(3K)$.


\subsection{Boundary Conditions and Initial state}
\label{VA}

When solving the Black-Scholes equation Eq. (\ref{BS}) after the change of variables $x= \log S$, the resulting equation, Eq. (\ref{BSlog}), turns out to be a partial differential equation with constant coefficients. This means that we can displace the initial condition by a given shift and solve the problem, in the sense that the actual solution of the original problem can be recovered afterwards by performing the same shift in the opposite direction. Indeed, given a bounded interval for the stock price $S\in [1/S_\text{max}, S_\text{max}]$, we make use of this property in order to have a symmetric initial condition with respect to $x$ as follows. We make a shift in order to translate the initial condition of the put option such that the support of this initial condition is the interval $[0,2 \log S_\text{max},]$, instead of $[-\log S_\text{max}, \log S_\text{max}]$, for a given discretization. After this shift, we make a duplication of the initial condition via a reflection in order to obtain periodic boundary conditions.

Moreover, due to time reversion, Black-Scholes initial value problem starts from the final payoff as initial condition. In terms of $x$, at maturity time, $\tau=0$, this condition results

\begin{equation}
    V_p(\tau=0, x)=\text{max}\{K-\exp (x),0\}.
\end{equation}

\begin{figure}[t]
\centering

\includegraphics[width=0.48\textwidth]{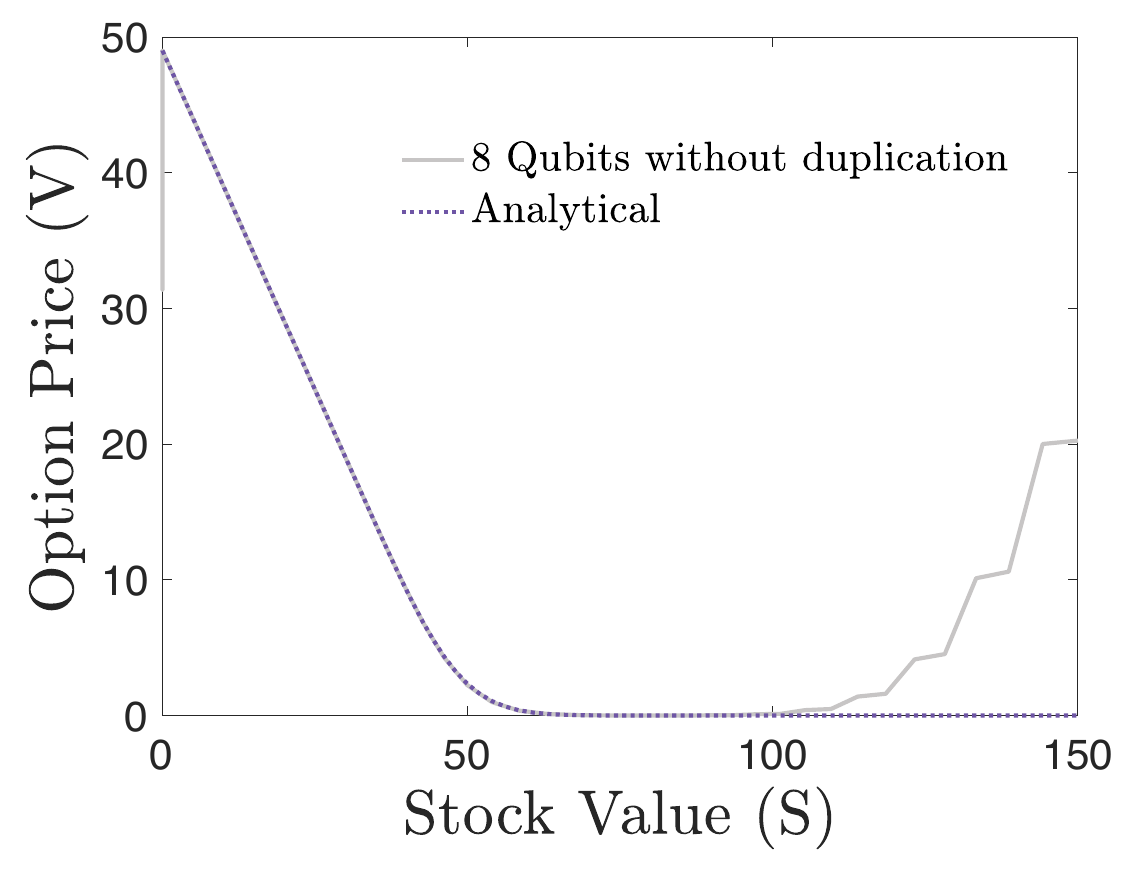}
\caption{ Solution to the Black-Scholes equation without duplicating the initial condition. We can appreciate how it presents a strong border effect due to periodic boundary conditions. Simulation parameters: $n=8$, $S_\text{max}=150 $ u, $K=50 $ u, $\sigma=0.2$, $r=0.04$, $T=1$ year. Scripts for numerical simulations can be found at \cite{github}}.
\label{sindup}
\end{figure}

Assuming that we use $n$ qubits to discretize the position space $x$, hence we have $N_x=2^{n}$  points, i.e. eigenstates $\ket{x_j}$, and each one of them corresponds to a discrete value of $x_j$.

\begin{equation}
   x_j= -x_\text{max}+\delta_xj \ \ \ j=0\hdots\ N_x-1,
\end{equation}
\begin{equation}
  \hat{X}\ket{x_j}= x_j\ket{x_j} \ \ \ j=0\hdots\  N_x-1,
\end{equation}
with $\delta_x=\frac{2x_\text{max}}{N_x-1}$. We use one of the $n$ qubits to duplicate the initial condition, which mitigates the border effects that would appear if we do not duplicate the initial condition when choosing periodic boundary conditions, see Fig. \ref{sindup}. In order to accomplish this duplication, we impose symmetry of the wave function with respect to $x=0$, hence the coefficients of the eigenstates $\ket{x_j} $ and $\ket{x_{N_x-1-j}}$ are the same $\forall  j=0\ ...\ N_x-1.$ Furthermore, this duplication reduces the size of the real price simulation space to the interval  $(-x_\text{max}/2, x_\text{max}/2)$, which is shifted to the interval $(- x_\text{max},0)$, as we pursue to duplicate the initial condition with respect to $x=0$. We calculate the value $N_\text{max}$ as the largest index $i$ such that $K-\exp (x_j)\geq 0$,
\begin{equation} 
K-\exp(x_{N_\text{max}})=0 \rightarrow  -\frac{x_\text{max}}{2}+\delta_xN_\text{max} =\log (K),
\end{equation}
thus
\begin{equation} 
N_\text{max}= \floor*{(N_x-1)\left(\frac{\log (K)}{2x_\text{max}}+\frac{1}{4}\right)}.
\end{equation}
Therefore, except for the normalization of the wave function, the coefficient of the eigenstate $\ket{x_j}$ is $K-\exp(-x_\text{max}/2+\delta_xj ),$ for $j=0\ ...\ N_\text{max}$. Considering that due to the duplication each coefficient is repeated twice, the norm squared results
\begin{equation}
\Lambda=\left(2\sum_{m=0}^{N_\text{max}}(K-\exp(-x_\text{max}/2+\delta_x m ))^2\right).
\label{normalization}
\end{equation}
Finally the normalized wave function reads
\begin{equation}
    \ket{V_p}=\sum_{j=0}^{N_\text{max}}\frac{K-e^{-x_\text{max}/2+\delta_x j}}{\Lambda^{1/2}}\left(\ket{x_j} + \ket{x_{N_x-1-j}}\right)
    \label{eqinitial}
\end{equation}
Moreover, we need an additional ancillary qubit, $q_E$, associated with the embedding, Eq. (\ref{Uemb}). Thus, the initial state of the embedded system results
$\ket{\Phi_0}=~\ket{0_E}\otimes \ket{V_p}$.

In the general case, loading an arbitrary state into a quantum computer requires an exponential quantity of gates \cite{UGD,DFB,TCR,SQL,PMB,ESP,SLQ}, which introduces a main drawback for an efficient simulation of the Hamiltonian dynamics.  Nevertheless, for some specific cases such as smooth differentiable functions, the initial state can be efficiently loaded into a gate-based quantum computer, as detailed in \cite{EIPD, QIMA, holmes2020efficient} . The work presented in Ref~\cite{EIPD} introduces two algorithms to achieve the efficient approximated loading of some functions, which in particular includes the initial state of the European Put options. 

\begin{table*}
\begin{tabular}{c|c|c}
Problem & Algorithm & Cost\\
\hline
\hline
Information Loading   & \ Amplitude encoding \cite{gonzalezconde2023efficient,EIPD} \ & \ \ \ $\mathcal{O}(\text{poly}(n)) $ \ \ \  \\
''  & Block Encoding \cite{Clader_2022} & \ \ \ $\mathcal{O}(\text{poly}(n)) $ \ \ \ \\
''  & Direct Classical Evaluation$^{*}$  & \ \ \ $\mathcal{O}(2^n) $ \ \ \ \\ \hline
PDE Simulation   & \ Quantum Signal Processing$^{**}$ \cite{Low_2017} \ & \ \ \ $\mathcal{O}\left( td\| \hat{H}\|_\mathrm{max} + \frac{\log(1/\epsilon)}{\log \log(1/\epsilon)}\right)$ \ \\
''  & Quantum Linear Solver$^{**}$ \cite{PRXQuantum.3.040303}  & \ \ \ $\mathcal{O}(s\kappa \log(1/\epsilon))$  \ \ \ \\
''  & Crank-Nicolson  \cite{QIMA,findif}& \ \ \ $\ \mathcal{O}(T_{\text{steps}}2^{n})$  \ \ \ \\
''  & Fast Fourier transform \cite{QIMA,findif}& \ \ \ $\mathcal{O}((n+1)2^{n})$ \ \ \ \\ 
''  & Matrix exponentiation \cite{matrixexp} & \ \ \ $\mathcal{O}(2^{n})$ \ \ \ \\ \hline
Expected Value   & \ Amplitude estimation \cite{Montanaro_2015, kubo2022pricing, Miyamoto_Kubo} \ & \ \ \ $\mathcal{O}(1/\epsilon) $ \ \ \  \\
''  & Classical Monte Carlo \cite{Montanaro_2015} & \ \ \ $\mathcal{O}(1/\epsilon^2) $ \ \ \ \\

\end{tabular}
\caption{Algorithms and their costs for solving the Black-Scholes PDE. We compare the costs of different tasks when working with multivariate functions, from the construction of the state, to the simulation of their evolution. Note that some classical methods also require the time discretization, which significantly contributes to the overall complexity. $\epsilon$, desired error bound in the particular task;  $n$, the number of qubits, equivalent to having $2^n$ degrees of freedom (points in the grid); and $T_{\text{steps}}$, the number of time steps.$^{*}$ Method only run one time $^{**}$ Query complexity.} 
\label{table1}
\end{table*}

\subsection{Efficient Simulation of the Black Scholes dynamics}

As the dynamics of Hamiltonian only depends on functions of the momentum operator $\hat{P}_x$, we can employ the quantum Fourier transform to simplify the implementation of the dynamics to the simulation of diagonal Hamiltonians, a special case of sparsity. Indeed,  we aim to simulate the operators $\hat{O}(\tau)$ and $e^{i\tau\hat{H}_{\text{BSH}}}$. By using Eq.(\ref{Fourier}) and the identity, $f\left(\hat{F}_{N_x}^{\dagger}\hat{P}_k \hat{F}_{N_x}\right)=\hat{F}_{N_x}^{\dagger}f \left( \hat{P}_k\right) \hat{F}_{N_x}$, where $f$ is an analytic function, the problem reduces itself to calculate the exponential of operator functions acting on diagonal momentum matrices.  In this way, an initial quantum Fourier transform on $N_x$ grid points, $\hat{F}_{N_x}$,  allows us to transform the initial condition encoded in the positions basis, Eq.~(\ref{eqinitial}), into the momentum space. After applying the diagonal operators, the inverse Fourier transform, $\hat{F}_{N_x}^{\dagger}$  enables us to recover the solution in position space, which encodes the price information. Note that the quantum Fourier transform is an efficient subroutine implemented with complexity $\mathcal{O}(n\log n)$. Therefore, we can assume from now on that the operators are diagonal, and consequently sparse.

Currently, quantum signal processing (QSP) techniques offer the most efficient quantum algorithms for quantum simulation of sparse Hamiltonians, as mentioned in \cite{Low_2017}. This is formally stated by the subsequent theorem, which establishes limits on the simulated time, accuracy, and success probability associated with QSP methods.

\begin{thm}[Optimal sparse Hamiltonian simulation using quantum signal processing (QSP) \cite{Low_2017}]\label{optsim}
A $d$-sparse Hamiltonian $\hat{H}$ on $n$ qubits with matrix elements specified to $m$ bits of precision can be simulated for time-interval $t$, error $\epsilon$, and success probability at least $1 - 2\epsilon$
with $\mathcal{O}\left( td\| \hat{H}\|_\mathrm{max} + \frac{\log(1/\epsilon)}{\log \log(1/\epsilon)}\right)$ oracle queries and a
factor $\mathcal{O}(n +~ m\, \mathrm{polylog}(m))$ additional quantum gates. The quantum simulation is valid for simulated time $td\| \hat{H}\|_\mathrm{max}  \sim ~ \mathcal{O}\left(\frac{\log(1/\epsilon)}{\log\log(1/\epsilon)}\right)$. 
\end{thm} 

Given that the Black Scholes Hamiltonian we have derived from the considered PDE is a fully quantum 1-sparse operator acting on a Hilbert space, we can conclude that simulations of the Black Scholes dynamics are optimal according to Theorem ~\ref{optsim}.  The oracles needed for this QSP methodology provide a description of the Hamiltonian, i.e.  where the sparse elements are (this is negligible as in this case the Hamiltonian can be diagonalized via QFT) and what is their value. In this sense, in order to build the oracle that returns the value of the Hamiltonian eigenvalues, which are analytically known, one might use the results in Ref. \cite{Wang_2020}. Additionally, in Tab.\ref{table1} we illustrate a comparison of complexities for solving the Black Scholes PDE with different methods.

\subsection{Measurement and Post selection}
\label{postselection}
\begin{figure*}[t]
  \centering
  \includegraphics[width=2\columnwidth]{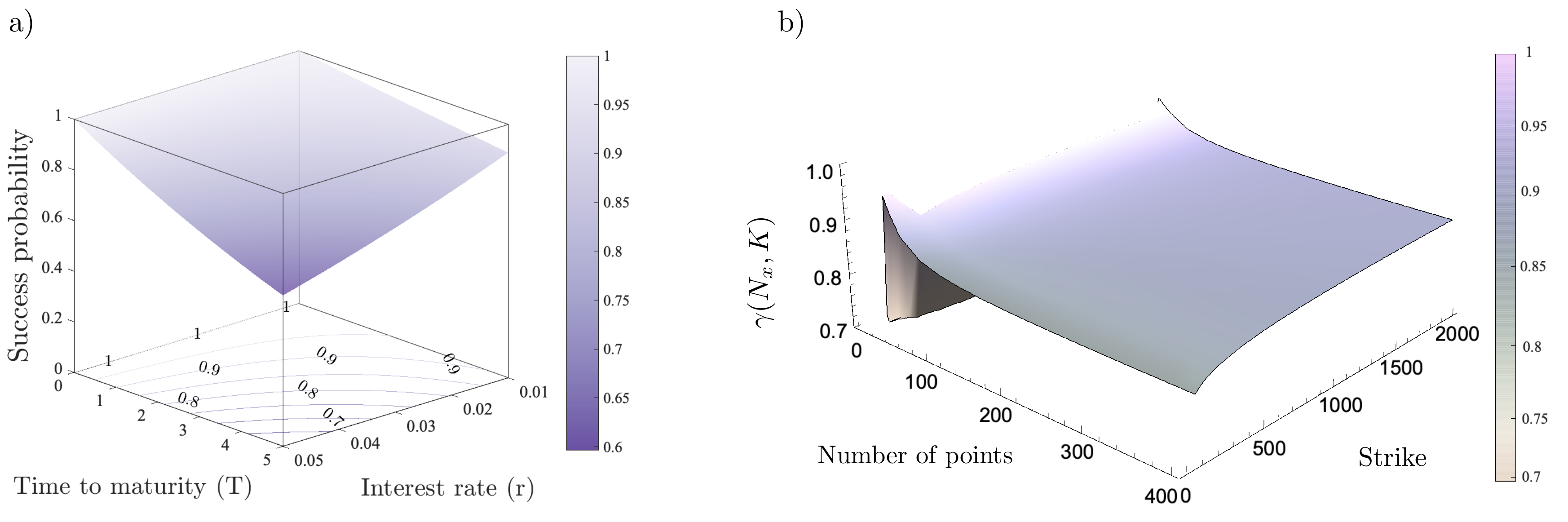}
  \caption{(a) Success probability in post-selection protocol corresponding to Eq.~(\ref{PS}) depending on time to maturity in years and risk-free interest rate. The probability is above 0.6 for all values in the mesh. (b) Lower bound probability of success $\gamma(N_x,K)$. As we can observe there exists an asymptotic convergence value for both, number of points and strike. The value of the asymptotic limit is over 0.6, what indicates that our protocol would be success in more than a half of the realizations. Parameters values: $S_{\text{max}}=150$ u, $K=50$ u, $\sigma=0.2$, $n=8$.  }
  \label{fig5}
\end{figure*}

Once the Hamiltonian dynamics have been efficiently simulated, the outcomes of the measurements in our circuit need to be postselected to recover the non-Hermitian dynamics of the Black-Scholes equation, which represents the option price. The first step is to measure the ancillary embedding qubit. If the measurement outcome is $\ket{0_E}$, we proceed to retrieve the price information. Otherwise, we discard the measurement. The probability of successfully recovering the desired dynamics, i.e., obtaining the value $\ket{0_E}$ when measuring the ancilla, depends on the expression

\begin{multline}
\label{PS}
Ps=\bra{V_p}\hat{F}_{N_x}^{\dagger}e^{-2T(\frac{\sigma^2}{2}\hat{P}_k^2+r\mathds{I})}\hat{F}_{N_x}\ket{V_p}\\ =\frac{1}{N_x \Lambda} \sum_{k=0}^{N_x-1}\Bigg[ \Bigg( \sum_{j=0}^{N_\text{max}} (K-e^{-x_\text{max}/2+j\delta_x})e^{2\pi i kj/N_x} +\\ \sum_{j=N_x-1-N_\text{max}}^{N_x-1} (K-e^{-x_\text{max}/2+(N_x-1-j)\delta_x})   e^{2\pi i kj/N_x}  \Bigg)\\ \Bigg( \sum_{j'=0}^{N_\text{max}} (K-e^{-x_\text{max}/2+j'\delta_x})e^{-2\pi i kj'/N_x} +\\ \sum_{j'=N_x-1-N_\text{max}}^{N_x-1} (K-e^{-x_\text{max}/2+(N_x-1-j')\delta_x})   e^{-2\pi i kj'/N_x} \Bigg) e^{-2T(\frac{\sigma^2}{2}p_k^2+r)} \Bigg]
\end{multline}
where $p_k$ is given by Eq. (\ref{eigen}). Considering that all the terms of the sum in $k$ are positive , the largest term corresponds to $k=0$. Thus, considering only this term, the success probability can be lower bounded by

\begin{multline}
P_s \geq \frac{1}{N_x \Lambda} e^{-2Tr} \Bigg[ \sum_{j,j'=0}^{N_\text{max}} (K-e^{-\delta_xN_x/4+j\delta_x}) (K-e^{-\delta_xN_x/4+j'\delta_x}) +\\ \sum_{j,j'=N-1-N_x}^{N_x-1} (K-e^{-\delta_xN_x/4+(N_x-1-j)\delta_x}) (K-e^{-\delta_xN_x/4+(N_x-1-j')\delta_x})    \\  \sum_{j=0,j'=N_x-1-N_\text{max}}^{N_\text{max},N_x-1} (K-e^{-\delta_xN_x/4+j\delta_x})(K-e^{-\delta_xN_x/4+(N_x-1-j')\delta_x})  +\\ \sum_{j=N_x-1-N_\text{max},j'=0}^{N_x-1,N_\text{max}} (K-e^{-\delta_xN_x/4+(N_x-1-j)\delta_x})(K-e^{-\delta_xN_x/4+j'\delta_x})   \Bigg].
\end{multline}
As the four terms of the expression above sum up to the same, we can define 

\begin{multline}
\small
\gamma_0 (N_x,K) =\sum_{j,j'=0}^{N_\text{max}} (K-e^{-x_\text{max}/2+j\delta_x}) (K-e^{-x_\text{max}/2+j'\delta_x})\\
= \frac{e^{-\delta_xN_x/2} \Bigg(1-e^{\delta_x(1+N_x)}+e^{\delta_x N_x/4}(-1+e^{\delta_x}K(1+N_\text{max}) \Bigg)^2}{(-1+e^{\delta_x})^2}
\end{multline}
and finally 
\begin{equation*}
\gamma (N_x,K)=\frac{4\gamma_0}{\Lambda N}.
\end{equation*}
As we can observe, the success probability, strongly depends on the maturity time and risk-free interest rate, but, for the usual range of financial parameters, its value is always above $0.6$ as depicted in  Fig.~\ref{fig5}~(a). Note that obtaining a probability of at least $1/2 + \epsilon$ with $\epsilon>0$ is a necessary ingredient for the successful deployment of the algorithm. The function $\gamma (N_x,K)$, depicted in Fig.~\ref{fig5}~(b), shows an asymptotic behaviour when $N _x\rightarrow \infty$,
\begin{equation}
\small{
    \lim_{N_x\rightarrow\infty} \gamma (N_x,K)= \frac{\left(-1+K^2-6K^2\log(K)\right)^2}{\left(-1+12K^2-11K^4+36K^4\log(K)\right)\log(3K)}.}
\end{equation}

If the system has evolved following the desired dynamic, we can retrieve the option price corresponding to the spot  $\ket{x_j}$, which encodes the stock price of interest $S_j=e^{x_j}$, by measuring the POVM$$ \bigg \{ \ket{x_j}\bra{x_j},  \mathds{I} - \ket{x_j}\bra{x_j} \bigg \}.$$ This task can be easily attained by using a multi-control gate acting on an extra qubit. In order to detail the proccess, let us suppose we have measured the state ancillary embedding qubit with probability $p(|0_E\rangle)\geq0.6$. Therefore the system has collapsed into a quantum state of the form $|\phi_f\rangle=~ \sqrt{1-a^2}|x_j^\perp\rangle\otimes|0_G\rangle  +~ a|x_j\rangle\otimes|0_G\rangle$ where $\langle x_j^\perp|x_j\rangle=0$ and $a=p(x_j \lvert  0_E)$  is the amplitude probability we desire to measure. If we apply a multi-control gate $U_{MCX}(|x_j\rangle)=~|x_j\rangle\langle x_j|~\otimes ~X +~ (I-~|x_j\rangle\langle x_j|)\otimes~ I$ acting on the ancillary qubit , we obtain the state $U_{MCX}(|x_j\rangle)|\phi_f\rangle=~ \sqrt{1-a^2}|x_j^\perp\rangle\otimes~|0_G\rangle  + a|x_j\rangle\otimes|1_G\rangle $. Consequently the estimation of  can be done by measuring the ancillary qubit in the computational basis.  If we consider a sampling process to retrieve the amplitude of a certain stock price and we want to determine with a precision error $\tilde{\epsilon}$, then we need  $\mathcal{O}(\frac{1}{\tilde{\epsilon}^2}) $ measurements, which can be quadratically improved by using the quantum amplitude estimation algorithm (QAE) by straightforwardly measuring the amplitude of $|x_j\rangle \otimes |0\rangle$  \cite{QAE, kubo2022pricing, Miyamoto_Kubo}. Therefore, assuming the constraint $e^{x_\text{max}/2}=3K$, it is possible to obtain a lower bound $Ps\geq  e^{-2Tr} \gamma (N_x,K)$. 

`We would like to remark that typically in a $n$ qubit state the amplitudes are of the order of $\sqrt{1/2^n}$ and therefore the number of rounds of QAE needed might result in an overhead of resources. A solution for this issue was proposed in Ref. \cite{Miyamoto_Kubo, kubo2022pricing}'.

Finally, the discrete value of solution for the Black-Scholes equation at maturity time, $T$, on the stock price $S_j=e^{x_j}$ is given by the expression

\begin{equation}
V_p(T, S_j)=\sqrt{p(x_j \cap 0_E)\Lambda},
\end{equation} 
where $p(x_j \cap 0_E)=p(x_j \lvert  0_E)p(0_E)\geq 0.6 p(x_j \lvert  0_E)$ is the probability of measuring the eigenstate $\ket{x_j}$ and the ancillary embedding qubit in the state $\ket{0_E}$, and $\Lambda$ is the normalization factor given by Eq.~(\ref{normalization}).

\section{Stock-price-dependent volatility}


Regarding the case of stock price-dependent volatility, we assume $\sigma=\sigma(\hat{x})$ and $\sigma(\hat{x})=\sigma^\dagger(\hat{x})$. Therefore, the Hamiltonian resulting from the Black-Scholes equation appears as $\hat{H}_{\text{BS}}=i\frac{\sigma^2(\hat{x})}{2}\hat{p}^2-\left(\frac{\sigma^2(\hat{x})}{2}-r \right)\hat{p}+ir\mathds{I}$ upon initial inspection. To achieve an appropriate quantization, it is necessary to include terms of the form $\frac{1}{2}\left(\sigma^2(\hat{x})\hat{P}+~\hat{P}\sigma^2(\hat{x})\right)$. Additionally, the dynamics of the entire Hamiltonian should be embedded, which significantly increases the complexity of the problem. The natural question that arises now is whether a choice of $\sigma(\hat{x})$ can maintain the sparsity of the Hamiltonian or if the system can be approximated by a controllable sparse Hamiltonian. It is worth noting that no analytical solution to the partial differential equation (PDE) exists in this case; hence, our algorithm would provide a meaningful numerical solution. Further investigation will be dedicated to studying this case in detail in subsequent works.

\section{Conclusions}
We have introduced a quantum algorithm to solve Black-Scholes partial differential equation in a digital quantum computer by mapping it to Schr\"odinger equation and the use Hamiltonian simulation techniques to simulate its dynamics. The non-Hermitian nature of the resulting Hamiltonian has been solved by embedding the dynamics into an enlarged Hilbert space, and by post-selecting the outcome of the simulation.  As a consequence of choosing periodic boundary conditions for the discretized momentum operator, and in order to improve the stability and performance of our algorithm, we also used a discretization qubit to duplicate the initial condition. Indeed, we have obtained a precision comparable to classical algorithms with a total of 9 qubits to simulate the Black-Scholes dynamics in a fault-tolerant quantum computer and an expected success probability value for the post-selection protocol above 60\%. Our perspective for a future work is to introduce errors associated to NISQ devices in order to analyze the realistic implementation in a near-term quantum platform. We want to highlight that the embedding techniques introduced may be extended to simulate the dynamics of the general non-Hermitian Hamiltonians and imaginary time evolution. This could allow us to introduce additional degrees of freedom in the model, e.g. spatial-time dependent volatility (stochastic local volatility) or coupled options. For instance, we could use the quantum principal component analysis raised in Ref. \cite{PCAS} together with coupled Black-Scholes models to address problems with coupled options. Moreover, the present work has been accomplished for the European option pricing problem, but it may be carried through simulate different kind of options, considering American and Asian options \cite{certo2023conditional}, for example.

\begin{acknowledgments}
We thank B. Candelas for useful discussions regarding the digital implementation. The authors acknowledge financial support from OpenSuperQ+100 (Grant No. 101113946) of the EU Flagship on Quantum Technologies, as well as from the EU FET-Open project EPIQUS (Grant No. 899368), also from Project Grant No. PID2021-125823NA-I00 595 and Spanish Ramón y Cajal Grant No. RYC-2020-030503-I funded by MCIN/AEI/10.13039/501100011033 and by “ERDF A way of making Europe” and “ERDF Invest in your Future”; this project has also received support from the Spanish Ministry of Economic Affairs and Digital Transformation through the QUANTUM ENIA project call - Quantum Spain, and by the EU through the Recovery, Transformation and Resilience Plan – NextGenerationEU within the framework of the Digital Spain 2026 Agenda. Authors also acknowledge funding from Basque Government through Grant No.  IT1470-22 and the IKUR Strategy under the collaboration agreement between Ikerbasque Foundation and BCAM on behalf of the Department of Education of the Basque Government, as well as from and UPV/EHU Ph.D. Grant No. PIF20/276.
\end{acknowledgments}

\end{document}